\documentclass[aps,prb,twocolumn,groupedaddress]{revtex4-2}

\usepackage[T1]{fontenc}
\usepackage{amsmath}
\usepackage{amssymb}
\usepackage{amsfonts}

\usepackage{bbm}
\usepackage{xspace}
\usepackage{color}
\usepackage{ulem} 
\usepackage{placeins}
\usepackage{bm}
\usepackage{siunitx}
\usepackage{nccmath}
\usepackage{mathtools} 
\usepackage{empheq}
\usepackage{enumerate}
\usepackage{graphicx}
\usepackage{dcolumn}
\usepackage[caption=false]{subfig}

\AddToHook{cmd/appendix/before}{%
  \setcounter{equation}{0}%
  \setcounter{theorem}{0}%
}

\usepackage[colorlinks=true,linkcolor=magenta,citecolor=blue,urlcolor=blue]{hyperref}

\usepackage{times}

\usepackage[version=4]{mhchem}

\graphicspath{{.}{Bilder/}}

\begin{document}

\title{Tunneling chirality Hall effect in type-I Weyl semimetals}
\author{W.~Zeng}
\email[E-mail: ]{zeng@ujs.edu.cn}
\affiliation{Department of physics, Jiangsu University, Zhenjiang 212013, China}


\begin{abstract}{}
We propose a tilt-assisted chirality Hall effect in the normal metal-superconductor (NS) junctions based on the time-reversal broken type-I Weyl semimetals. It is found that the chirality-contrasting skew reflection occurs at the NS interface due to the tilt of the Weyl cones, which is responsible for the nonzero transverse chirality Hall currents. Distinct from the Hall effect induced by the Berry curvature, we further illustrate that the transverse chirality current here is determined by the symmetry of the tilt. Specifically, both the transverse chirality Hall current and the transverse charge Hall current may occur when the tilt breaks the mirror symmetry ($\mathcal{M}$). However, a pure transverse chirality Hall current with zero net charge is present when the tilt breaks $\mathcal{M}$ symmetry but preserves the combined $\mathcal{MC}$ symmetry, where $\mathcal{C}$ represents the $\mathbb{Z}_2$ exchange symmetry.



\end{abstract}

\maketitle

\section{Introduction}\label{sec:1}

Weyl semimetals are three-dimensional topological materials with the conduction and valence bands touching at two or more crossing points in the bulk, which are known as the Weyl nodes \cite{RevModPhys.90.015001,PhysRevB.83.205101,PhysRevB.95.041104}. The emergence of the nontrivial and stable Weyl
nodes requires the breaking of either time-reversal symmetry ($\mathcal{T}$) or spatial inversion symmetry ($\mathcal{I}$). As a consequence of the Nielsen-Ninomiya theorem \cite{nielsen1981absence}, the minimal model of the $\mathcal{T}$ symmetry broken Weyl semimetal contains a single pair of Weyl nodes, whereas the $\mathcal{I}$ symmetry broken one contains four Weyl nodes \cite{PhysRevLett.107.127205,PhysRevX.5.011029}. Each of the paired Weyl nodes acts like a topological charge with the charge sign corresponding to its chirality \cite{chen2022discovery,chang2018topological}. The manipulation of the chirality is one of the hot topics in Weyl physics. Up to now, several works have been devoted to such chirality-dependent physics of Weyl semimetals, such as the chiral anomaly \cite{PhysRevB.86.115133,burkov2015chiral,xiong2015evidence}, chirality-dependent Hall effect \cite{PhysRevLett.115.156603,PhysRevB.102.121105,PhysRevLett.126.236601} and chirality Josephson effect \cite{PhysRevLett.121.226604,PhysRevB.103.125147}.





A finite tilt of the Weyl cones can be generated because the Lorentz symmetry is not necessarily the symmetry group in condensed matter systems \cite{PhysRevB.100.045144}. Distinct from the graphene-like materials where the tilting is usually weak \cite{PhysRevB.78.045415,PhysRevB.88.045126,mao2011graphene}, the tilting can be strong in Weyl semimetals. Depending on whether the Weyl cone is overtilted or not, Weyl semimetals can be classified into two subgroups, \textit{i.e.}, type-I and type-II Weyl semimetals \cite{li2017evidence,soluyanov2015type,PhysRevX.5.031013,xu2015discovery}. The type-I Weyl semimetals possess a closed Fermi surface enclosing either an electron or a hole pocket, with a vanishing density of states at the Weyl nodes. The type-II Weyl semimetals host overtilted Weyl cones and the Fermi surface near the Weyl nodes is hyperboloidal with a large density of states, leading to the electron and hole pockets near the Weyl nodes \cite{PhysRevLett.117.077202,ma2019nonlinear}. Many tilt-induced intriguing transport properties have been reported in Weyl semimetals. Such as the anomalous Nernst and thermal Hall effects \cite{PhysRevB.96.115202}, double Andreev reflection \cite{PhysRevB.96.155305}, tilt-assisted $\pi$-phase Josephson current \cite{PhysRevB.102.085144}, and linear magnetochiral transport \cite{PhysRevB.99.085405}.




A very recent interesting work \cite{PhysRevLett.131.246301} reported that the tilt mechanism can lead to the tunneling valley Hall effect in Dirac systems, where a strong tilt-dependent transverse valley Hall current can be generated by the momentum filtering of the tunneling Dirac fermions. Subsequently, the nonlinear valley Hall effect in tilted massless Dirac fermions in strained graphene and organic semiconductors was predicted~\cite{PhysRevLett.132.096302}, where a valley Hall current occurs with both inversion and time-reversal symmetry. Inspired by this, here we propose a tilt-induced chirality Hall effect in
Weyl semimetals. We theoretically investigate the transverse charge and chirality transport in the normal metal-superconductor (NS) junctions based on the type-I Weyl semimetals, which breaks the time-reversal symmetry but preserves the inversion symmetry. It is found that the scattering at the NS interface is chirality-dependent. For the electrons with a given chirality, the incident angle resolved reflection probability is asymmetric, which is responsible for the transverse chirality Hall current. The chirality Hall conductance arsing from the skew scattering can be obtained within the Landauer formalism \cite{datta1997electronic}. It is found that the transverse chirality current is determined by the tilt parameter $c$ and the intersection angle $\alpha$ between the line connecting the two opposite chiral Weyl nodes and the normal of the NS interface. For the tilt breaking the mirror symmetry $\mathcal{M}$, which requires $\alpha\neq\pi/2$, both the transverse chirality Hall current and charge Hall current may occur. However, for the tilt breaking the $\mathcal{M}$ symmetry but preserving the combined $\mathcal{MC}$ symmetry with $\mathcal{C}$ being the $\mathbb{Z}_2$ exchange symmetry, requiring $\alpha=0$ or $\pi$, a pure transverse chirality Hall current with zero net charge appears.

The remainder of the paper is organized as follows. The model Hamiltonian and the scattering approach are explained in detail in Sec. \ref{sec:2}. The numerical results and discussions are presented in Sec. \ref{sec:3}. Finally, we conclude in Sec. \ref{sec:4}.

\section{Model}\label{sec:2}
\begin{figure}[!tp]
\centerline{\includegraphics[width=1\linewidth]{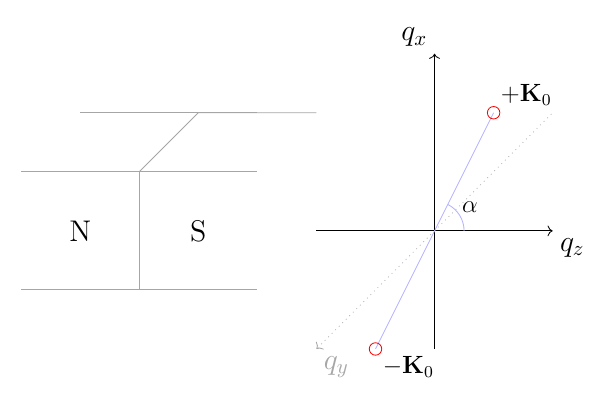}}
\caption{\label{fig:1}
(Left) Sketch of the NS junction under consideration, where the normal of the NS interface is along the $z$ axis. (Right) Schematic of the momentum space with two Weyl nodes at momenta $\pm\mathbf{K}_0$. The angle between two Weyl nodes and the $q_z$ axis is denoted by $\alpha$.
}
\end{figure}

We consider the NS junction along the $z$ axis, where the normal and superconduting regions are located at $z<0$ and $z>0$, respectively, as shown in Fig.\ \ref{fig:1}. The Weyl nodes $\pm\mathbf K_0$ are in the $q_x-q_z$ plane and the line connecting $\pm \mathbf K_0$ can make an angle of $\alpha$ with $q_z$ axis. In the crystal coordinates, the minimal model for the tilted Weyl semimetal is described by the effective two-band Hamiltonian \cite{chen2013specular,PhysRevB.95.064511}
\begin{align}
    H=\sum_{\chi,\mathbf{q}}\Psi^\dagger_{\chi,\mathbf q} \mathcal{H}_\chi(\mathbf q)\Psi_{\chi,\mathbf q},
\end{align}
where $\Psi_{\chi,\mathbf q}=(\psi_{\chi,\mathbf q\uparrow},\psi_{\chi,\mathbf q\downarrow})^T$ is the spinor basis with $\chi=\pm$ being the chirality of the Weyl nodes and $\mathbf q=(q_1,q_2,q_3)$ is the momentum measured from $\chi\mathbf K_0$. Around the Weyl nodes, the low-energy Hamiltonian reads \cite{PhysRevB.102.085144,PhysRevB.99.174501}
\begin{align}
\mathcal{H}_\chi(\mathbf{q})=&\hbar v_\chi q_1\sigma_0+\hbar v_F(q_1\sigma_1+q_2\sigma_2-\chi q_3\sigma_3),\label{Eq:hw}
\end{align} 
where $v_F$ is the Fermi velocity, $\sigma_0$ is the identity matrix and $\sigma_{i}\ (i=1,2,3)$ are Pauli matrices acting on the spin space. The tilt of the Weyl cones is along the $q_1$ direction with the parameter $v_\chi$. Here we focus on the tilting effect in the type-I Weyl semimetals, \textit{i.e.}, $|v_\chi|<v_F$. The inversion symmetry of the Hamiltonian in Eq.\ (\ref{Eq:hw}) requires $\sigma_3\mathcal{H}_+(\mathbf{q})\sigma_3=\mathcal{H}_-(-\mathbf{q})$, leading to $v_+=-v_-$, which implies that the opposite chiral Weyl cones have tilts in opposite directions. It is convenient to work with the junction coordinates, where the transport direction is assumed to be along the $z$ axis, as shown in Fig.\ \ref{fig:1}. The two different coordinate systems are related by the rotation transformation
\begin{align}
q_1=&q_x\cos\alpha-q_z\sin\alpha,\quad q_2=q_y,\nonumber\\
q_3=&q_z\cos\alpha+q_x\sin\alpha,
\end{align}
and, similarly, $\sigma_1=\sigma_x\cos\alpha-\sigma_z\sin\alpha$, $\sigma_2=\sigma_y$, $\sigma_3=\sigma_z\cos\alpha+\sigma_x\sin\alpha$, where $\alpha$ is the angle between the line connecting two Weyl nodes and the $q_z$ axis.

In the superconducting region, the zero-momentum BCS pairing is preferred for inversion-symmetric Weyl semimetals \cite{PhysRevB.92.035153}, for which the paired electrons are from two Weyl nodes with opposite chirality. This BCS superconductivity can be induced by the conventional superconductor via the proximity effect in the Weyl semimetal based junction \cite{volkov1995proximity,chen2013specular}. The pairing Hamiltonian reads
\begin{align}
\mathcal{H}_\Delta=\sum_{\chi,s}\int d\mathbf{r}\ \Delta\psi^\dagger_{\chi,s}(\mathbf{r})\psi^\dagger_{-\chi,-s}(\mathbf{r})+h.c.,
\end{align}
where $s=\{\uparrow,\downarrow\}$ is the spin index and $\Delta$ is the pairing potential. In the Nambu basis $(\psi_{\chi\uparrow},\psi_{\chi\downarrow},\psi_{-\chi\downarrow}^\dagger,-\psi_{-\chi\uparrow}^\dagger)^T$, the NS junction is described by the Bogoliubov-de Gennes (BdG) Hamiltonian \cite{PhysRevLett.97.067007,PhysRevB.106.094503,de2018superconductivity}
\begin{widetext} 
\begin{align}
\mathcal{H}_{BdG}=&\begin{pmatrix}
\mathcal{H}_\chi(-i\bm\nabla-\chi\mathbf{K}_0)-\mu(z)&\Delta(z)\\
\Delta^*(z)&\mu(z)-\sigma_y\mathcal{H}^*_{-\chi}(-i\bm\nabla+\chi\mathbf{K}_0)\sigma_y
\end{pmatrix},\label{Eq:bdg}
\end{align}
where the chemical potential $\mu(z)=\mu$ for $z<0$ and $\mu(z)=\mu_s$ for $z>0$, the pairing term $\Delta(z)=0$ for $z<0$ and $\Delta(z)=\Delta_0$ for $z>0$. By performing the translation transformation $D=\exp{(i\chi\mathbf{K}_0\cdot\mathbf{r})}$ and the spin rotation transformation $U=\sigma_x\exp(i\alpha\sigma_y)$ \cite{PhysRevLett.121.226604,PhysRevB.102.085144}, Eq.\ (\ref{Eq:bdg}) can be written as
\begin{align}
\mathcal{H}_{BdG}=&\begin{pmatrix}
\hbar v_F\big(\chi c q_t\sigma_0+(q_x\sigma_x-\chi q_y\sigma_y+ q_z\sigma_z)\big)-\mu(z)&\Delta_B(z)\\
\Delta_B(z)&\hbar v_F\big(-\chi c q_t\sigma_0-(q_x\sigma_x+\chi q_y\sigma_y+q_z\sigma_z)\big)+\mu(z)
\end{pmatrix},
\end{align}
where $c=|v_\chi/v_F|$ is a dimensionless parameter employed here to characterize the tilt ($0<c<1$), $q_t=q_x\cos\alpha-q_z\sin\alpha $ and $\Delta_B(z)=-\Delta(z)\sin\alpha\sigma_z+\Delta(z)\cos\alpha\sigma_x$.
\end{widetext}

In the normal segment of the junction ($z<0$), the scattering states propagating along the $+z$ axis are given by
\begin{align}
|\varphi_{e}^{>}\rangle=\begin{pmatrix}
\Gamma_{e}^>\\|\mathbf q_\parallel|e^{-i\phi}\\0\\0
\end{pmatrix}e^{iq_{e}^> z},\quad
|\varphi_{h}^>\rangle=\begin{pmatrix}
0\\
0\\
\Gamma_h^>\\
|\mathbf q_\parallel|e^{i\phi}
\end{pmatrix}e^{iq_{h}^> z},\label{Eq:wave}
\end{align}
where the subscripts `$e/h$' of the wave functions denote the electron/hole states, respectively, $\mathbf q_\parallel=(q_x,q_y)$ is the conserved transverse wave vector, $\phi=\arctan(\chi q_y/q_x)$, and $\Gamma_{e(h)}^>=(1+\chi c\sin\alpha)q_{e(h)}^>+(-)p_{e(h)}$ with $p_{e(h)}=(E+(-)\mu)/\hbar v_F-(+)\chi q_x c \cos\alpha$. We note that the factor $e^{i\mathbf{q}_\parallel\cdot\mathbf{r}_\parallel}$ with $\mathbf r=(x,y)$ is omitted in Eq.\ (\ref{Eq:wave}) for simplicity.
The longitudinal wave vectors are given by   
\begin{gather}
q_{e(h)}^>=\frac{+(-)\chi p_{e(h)}c \sin\alpha+\zeta_{e(h)}\sqrt{p_{e(h)}^2-\kappa|\mathbf q_\parallel|^2}}{\kappa},\label{Eq:vec}
\end{gather}
where
\begin{gather}
\zeta_{e(h)}=\mathrm{sgn}(p_{e(h)}+\sqrt{\kappa}|\mathbf q_\parallel|),\label{Eq:ztt}
\end{gather}
with $\kappa=1-(c\sin\alpha)^2$.
Similarly, the scattering states propagating along the $-z$ axis, \textit{i.e.}, $|\varphi^{<}\rangle$, can be obtained by the replacement $\zeta_{e(h)}\rightarrow-\zeta_{e(h)}$ in Eqs.\ (\ref{Eq:wave}-\ref{Eq:ztt}).

In the heavily doping limit $|\mu_s|\gg|\mu|$ \cite{RevModPhys.80.1337,PhysRevB.104.075436}, only the excitations quasiperpendicular transmitting to the superconducting region need to be considered, resulting in the effective excitation gap $\Delta_B=\Delta_0|\sin\alpha|$ \cite{PhysRevB.95.064511,PhysRevB.99.174501}. The transmitted states in the superconducting region are given by
\begin{align}
|\varphi_{s,1}^{>}\rangle=\begin{pmatrix}
    \mathcal{P}\\
    0\\
    -\Delta_B\\
    0
\end{pmatrix}e^{iq_{s,1}^{>}z},\quad |\varphi_{s,2}^{>}\rangle=\begin{pmatrix}
    0\\
    \Delta_B\\
    0\\
    \mathcal{P}
\end{pmatrix}e^{iq_{s,2}^{>}z},
\end{align}
where $\mathcal{P}=\sqrt{E^2-\Delta_B^2}+E$ and the subscript `$1(2)$' denotes the electron-like (hole-like) quasiparticle state. The longitudinal wave vectors in the superconducting region are given by
\begin{align}
q_{s,1(2)}^{>}=(\hbar v_F)^{-1}\frac{+(-)\mu_s+\sqrt{E^2-\Delta_B^2}}{1-(+)\chi c\sin\alpha}.
\end{align}

The total wave function describing the scattering process reads
\begin{align}
\psi(z)=\begin{cases}
|\varphi_{e}^{>}\rangle+r|\varphi_{e}^{<}\rangle+r_{A}|\varphi_{h}^{<}\rangle,& z<0, \\ 
t_1|\varphi_{s,1}^{>}\rangle+t_2|\varphi_{s,2}^{>}\rangle, &z>0.
\end{cases}
\end{align}
Here $t_{1(2)}$, $r$ and $r_{A}$ are the transmission amplitude, normal reflection amplitude and Andreev reflection amplitude, respectively, which can be obtained by matching the wave function at $z=0$. 

The longitudinal conductance for the $\chi$ chirality Weyl node is given by the Blonder-Tinkham-Klapwijk approach \cite{PhysRevLett.115.056602}
\begin{align}
\sigma_{z,z}^\chi=\frac{e^2}{h}\sum_{\mathbf{q}_\parallel}\Big(1+R_A-R\Big),
\end{align}
where
\begin{align}
R_A=\left|\frac{v_{h,z}^<}{v_{e,z}^>}\right||r_{A}|^2,\quad R=\left|\frac{v_{e,z}^<}{v_{e,z}^>}\right||r|^2,\label{Eq:AB}
\end{align}
are the Andreev reflection probability and normal reflection probability, respectively. We note that both $r$ and $r_A$ are the function of $\chi$. $v_{\varsigma,i}^\varrho$ ($\varsigma=e,h$, $\varrho=>,<$ and $i=x,y,z$) in Eq.\ (\ref{Eq:AB}) is the group velocity along the $i$ axis for the excitation state $|\varphi^\varrho_\varsigma\rangle$, which can be obtained by the Hellmann-Feynman theorem \cite{PhysRev.56.340}:
\begin{align}
v_{\varsigma,i}^\varrho=&\frac{\partial E_\varsigma}{\hbar\partial q_{i}^\varrho}=\frac{\partial}{\hbar\partial q_{i}}\langle\varphi^\varrho_\varsigma|\mathcal{H}_{BdG}|\varphi^\varrho_\varsigma\rangle\nonumber\\
=&\hbar^{-1}\langle\varphi^\varrho_\varsigma|\frac{\partial}{\partial q_{i}}\mathcal{H}_{BdG}|\varphi^\varrho_\varsigma\rangle\nonumber\\
=&\hbar^{-1}\langle\varphi^\varrho_\varsigma|\hat{j}_i|\varphi^\varrho_\varsigma\rangle,
\end{align}
with $\hat{j}_x=\tau_z(\sigma_x+\chi c\cos\alpha)$, $\hat{j}_y=-\chi\tau_z\sigma_y$ and $\hat{j}_z=\tau_z(\sigma_z-\chi c\sin\alpha)$.

The transverse conductance can be calculated within the Landauer formalism \cite{datta1997electronic,PhysRevB.24.2978,bagwell1989landauer} (see Appendix for details)
\begin{align}
\sigma_{\eta,z}^\chi=\frac{e^2}{h}\sum_{\mathbf{q}_\parallel}\Big(\frac{v_{e,\eta}^>}{v_{e,z}^>}-\frac{v_{e,\eta}^<}{v_{e,z}^>}|r|^2+\frac{v_{h,\eta}^<}{v_{e,z}^>}|r_{A}|^2\Big),\label{Eq:app1}
\end{align}
where $\eta=x,y$. The charge Hall angle ($\vartheta$) and the chirality Hall angle ($\vartheta_{\mathrm{chi}}$) are given by
\begin{align}
    \tan(\vartheta)=\frac{\sum_\chi \sigma_{\eta,z}^\chi}{\sum_\chi\sigma_{z,z}^\chi},\quad  \tan(\vartheta_{\mathrm{chi}})=\frac{\sum_\chi \chi\sigma_{\eta,z}^\chi}{\sum_\chi\sigma_{z,z}^\chi}.\label{Eq:app2}
\end{align}

\begin{figure}[!tp]
\centerline{\includegraphics[width=1\linewidth]{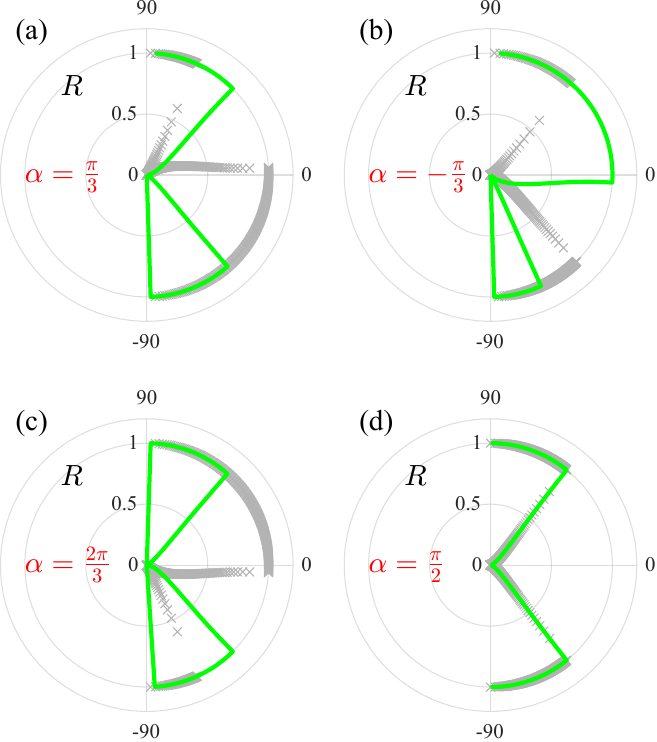}}
\caption{\label{fig:2}
The reflection probability $R$ versus the incident angle $\theta_i$ for $\chi=+1$ (gray) and $\chi=-1$ (green) chiral nodes. Panels (a-d) correspond to $\alpha=\pi/3$, $\alpha=-\pi/3$, $\alpha=2\pi/3$, and $\alpha=\pi/2$, respectively. The other parameters are $c=0.6$, $E=0.3\Delta_0$, and $\mu=1.2\Delta_0$. 
}
\end{figure}

\begin{figure}[!tp]
\centerline{\includegraphics[width=1\linewidth]{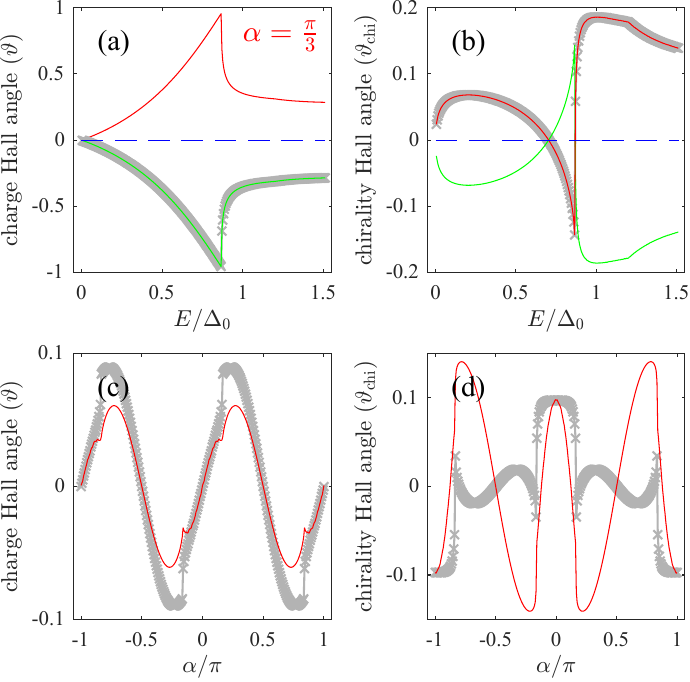}}
\caption{\label{fig:3}
[(a), (b)] Charge and chirality Hall angle as a function of the incident energy $E$ for $\alpha=\pi/3$ (red), $\alpha=-\pi/3$ (gray), $\alpha=2\pi/3$ (green), and $\alpha=\pi/2$ (blue dashed). The other parameters are $c=0.6$ and $\mu=1.2\Delta_0$. [(c), (d)] Charge and chirality Hall angle as a function of $\alpha$ for $\mu=1.2\Delta_0$ (gray) and $\mu=0.2\Delta_0$ (red).
}
\end{figure}

\section{Results}\label{sec:3}

\subsection{Skew reflection and transverse chirality and charge currents} 
\label{sub:1}

We first consider the situation where the effective superconducting gap $\Delta_B=\Delta_0|\sin\alpha|$ is nonzero, \textit{i.e.}, $\alpha\neq0,\pi$. Both the Andreev process and the normal reflection process contribute to the transverse Hall currents. It is sufficient to consider the normal reflection probability $R$ in the subgap regime ($|E|<\Delta_0|\sin\alpha|$) on account of $R+R_A=1$. 

The normal reflection probability $R$ versus the incident angle $\theta_i$ is shown in Fig.\ (\ref{fig:2})(a) for $\alpha=\pi/3$, where the effective superconducting gap is $\Delta_B=0.86\Delta_0$. Due to the Weyl cones tilt in the $q_x-q_z$ plane in our model, we focus on the reflection in the $x-z$ plane, where the incident angle is given by
\begin{align}
\theta_i=\arctan\left(\frac{v^>_{e,x}}{v^>_{e,z}}\right).
\end{align}
It is shown that the electrons from the $\chi=+1$ chiral node have large reflection probabilities for $-90^{\circ}<\theta_i<0^{\circ}$ [gray line in Fig.\ (\ref{fig:2})(a)], which is responsible for the nonzero transverse chirality Hall current. For the electrons from the $\chi=-1$ chiral node, the skew reflection also exists [green line in Fig.\ (\ref{fig:2})(a)]. However, the scattering is asymmetric for the electrons from different chiral nodes, \textit{i.e.}, $R_{\chi,\alpha=\pi/3}(\theta_i)\neq R_{-\chi,\alpha=\pi/3}(-\theta_i)$, indicating the presence of a nonzero transverse charge Hall current.

The other two scenarios with $\alpha=-\pi/3$ and $\alpha=2\pi/3$ are also considered, where the effective superconducting gaps are both $0.86\Delta_0$ as well. The normal reflection probabilities for $\alpha=-\pi/3$ and $\alpha=2\pi/3$ are shown in Figs.\ (\ref{fig:2})(b) and (\ref{fig:2})(c), respectively. The chirality-contrasting reflection remains present. However, compared with the $\alpha=\pi/3$ case, the electrons with the opposite (same) chirality are skew reflected to the opposite direction for $\alpha=-\pi/3$ ($2\pi/3$). Consequently, the scattering patterns in Figs.\ (\ref{fig:2})(a) and (\ref{fig:2})(b) are symmetric for different chiralities whereas the scattering patterns in Figs.\ (\ref{fig:2})(a) and (\ref{fig:2})(c) are symmetric for the same chirality. The above symmetric relations can be expressed in terms of $R$, namely 
\begin{gather}
R_{\chi,\alpha=\frac{\pi}{3}}(\theta_i)=R_{-\chi,\alpha=-\frac{\pi}{3}}(-\theta_i),\label{Eq:aw1}\\
R_{\chi,\alpha=\frac{\pi}{3}}(\theta_i)=R_{\chi,\alpha=\frac{2\pi}{3}}(-\theta_i).\label{Eq:aw2}
\end{gather}
For $\alpha=\pi/2$, the skew reflection is absent, and the scattering pattern for $\chi=+1$ and $\chi=-1$ are identical to each other , as shown in Fig.\ (\ref{fig:2})(d).

This chirality-contrasting skew reflection mentioned above may result in a transverse chirality Hall current as well as a transverse charge Hall current, which can be characterized by the Hall angle. The charge and chirality Hall angles versus the incident energy at different $\alpha$ are shown in Figs.\ (\ref{fig:3})(a) and (\ref{fig:3})(b), respectively. For $\alpha=\pi/3$, the charge Hall angle $\vartheta$ is positive and increases with the increasing of $E$ in the subgap energy regime ($|E|<0.86\Delta_0$), as the shown in Fig.\ (\ref{fig:3})(a) (red solid line). However, the chirality Hall angel is not monotonically dependent on $E$ in the subgap regime, as the shown in Fig.\ (\ref{fig:3})(b) (red solid line). For $\alpha=-\pi/3$ and $2\pi/3$, the charge Hall angles are equal to each other but negative, as the gray and green lines shown in Fig.\ (\ref{fig:3})(a), respectively. The chirality Hall angle remains unchanged for $\alpha=-\pi/3$ [gray line in Fig.\ (\ref{fig:3})(b)], but reverses its sign for $\alpha=2\pi/3$ [green line in Fig.\ (\ref{fig:3})(b)]. For $\alpha=\pi/2$, both the charge and chirality Hall angle are zero due to the absence of the skew reflection, as shown in Figs.\ (\ref{fig:3})(a) and (\ref{fig:3})(b) (blue dashed lines).

The $\alpha$-dependence of the Hall angle is shown in Figs.\ (\ref{fig:3})(c) and (\ref{fig:3})(d). The charge Hall angle $\vartheta(\alpha)$ is odd parity, as shown in Figs.\ (\ref{fig:3})(c), whereas the chirality Hall angle $\vartheta_{\mathrm{chi}}(\alpha)$ is even parity, as shown in Figs.\ (\ref{fig:3})(d). It is shown that both $\vartheta$ and $\vartheta_{\mathrm{chi}}$ are absent at $\alpha=\pm\pi/2$. However, for $\alpha=0$ and $\pi$, $\vartheta_{\mathrm{chi}}$ is finite but $\vartheta=0$.


The different $\alpha$-dependence of $\vartheta$ and $\vartheta_{\mathrm{chi}}$ can be understood by analyzing the reflection probability $R$. In fact, the symmetric scattering behaviors expressed in Eqs.\ (\ref{Eq:aw1}-\ref{Eq:aw2}) are valid for arbitrary parameter set $\{\alpha,-\alpha,\pi-\alpha\}$, which are given by
\begin{gather}
R_{\chi,\alpha}(\theta_i)=R_{-\chi,-\alpha}(-\theta_i),\label{Eq:sy1}\\
R_{\chi,\alpha}(\theta_i)=R_{\chi,\pi-\alpha}(-\theta_i).\label{Eq:sy2}
\end{gather}
For the Andreev reflection probability $R_{A}$, the similar identities can be obtained by using the current conservation relation $R_A=1-R$. With the help of Eqs.\ (\ref{Eq:app1}-\ref{Eq:app2}), the $\alpha$-dependent charge and chirality Hall angles can be expressed in terms of $R$, which are given by
\begin{gather}
\vartheta(\alpha)=\mathcal{Q}\sum_\chi\int_{-\frac{\pi}{2}}^{\frac{\pi}{2}}d\theta_i\sin\theta_iR_\chi(\theta_i,\alpha),\label{Eq:sx1}\\
\vartheta_{\mathrm{chi}}(\alpha)=\mathcal{Q}\sum_\chi\int_{-\frac{\pi}{2}}^{\frac{\pi}{2}}d\theta_i\chi \sin\theta_iR_\chi(\theta_i,\alpha),\label{Eq:sx2}
\end{gather}
with $\mathcal{Q}$ being a parameter independent of $\theta_i$ and $\mathcal{Q}(\alpha)=\mathcal{Q}(-\alpha)$. By substituting Eqs.\ (\ref{Eq:sy1}-\ref{Eq:sy2}) into Eqs.\ (\ref{Eq:sx1}-\ref{Eq:sx2}) and changing the integration variable $\theta_i\rightarrow-\theta_i$, one finds
\begin{gather}
\vartheta(\alpha)=-\vartheta(-\alpha)=-\vartheta(\pi-\alpha),\label{Eq:th1}\\
\vartheta_{\mathrm{chi}}(\alpha)=\vartheta_{\mathrm{chi}}(-\alpha)=-\vartheta_{\mathrm{chi}}(\pi-\alpha),\label{Eq:th2}
\end{gather}
which indicates that the odd-parity charge Hall angle $\vartheta(\alpha)$ disappears at $\alpha=\{\pi/2,-\pi/2,0,\pi\}$ and the even-parity chirality Hall angle $\vartheta_{\mathrm{chi}}(\alpha)$ disappears at $\alpha=\{\pi/2,-\pi/2\}$.

\subsection{Pure transverse chirality currents} 
\label{sub:2}

It is noted that the charge Hall angle $\vartheta$ vanishes whereas the chirality Hall angle $\vartheta_\mathrm{chi}$ remains finite at $\alpha=\{0,\pi\}$ [see Figs.\ (\ref{fig:3})(c) and (\ref{fig:3})(d)], implying a pure chirality current. 

\begin{figure}[!tp]
\centerline{\includegraphics[width=1\linewidth]{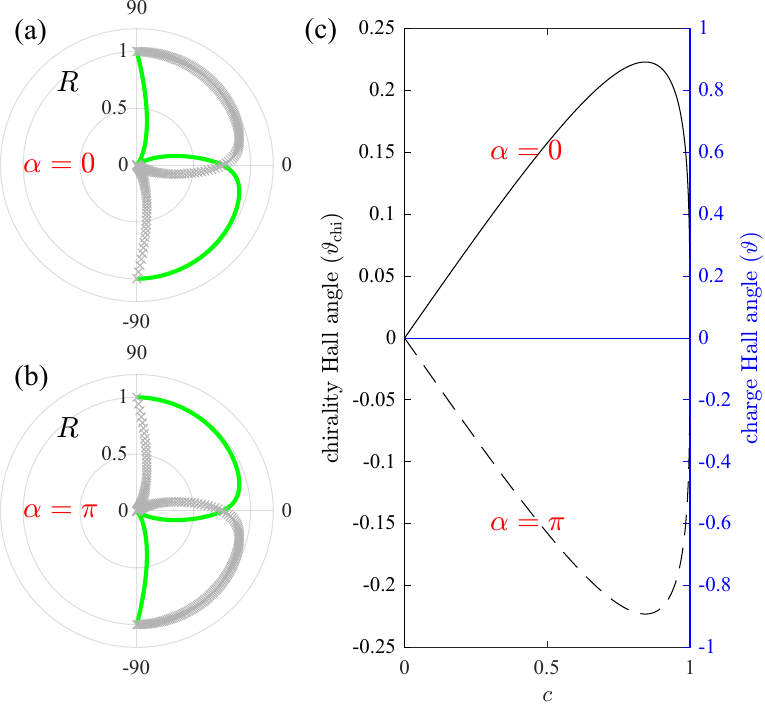}}
\caption{\label{fig:4}
[(a), (b)] The reflection probability $R$ versus the incident angle $\theta_i$ for $\chi=+1$ (gray) and $\chi=-1$ (green) chiral nodes. (c) Hall angle as a function of $c$.
}
\end{figure}

In this scenario the effective pairing potential is zero. Under the heavily doping condition, the incident electron is local reflected with probability $R$ in the normal region and quasiperpendicular transmitted ($\mathbf q_\parallel=0$) as an electron-like quasiparticle with probability $T_1=|v_{s1,z}^>/v_{e,z}^>||t_1|^2$ in the superconducting region. The current conservation requires $R+T_1=1$. In the heavily doping limit $|\mu_s|\gg|\mu|$, the normal reflection probability is only determined by the tilting parameter $c$. For $\alpha=0$, the normal reflection probability is obtained analytically as
\begin{align}
R=\left|\frac{\sqrt{\sec^2\theta_i-c^2}-1}{\tan\theta_i-\chi c}\right|^2,\label{Eq:rr}
\end{align}
where $\theta_i=\arctan(v^>_{e,x}/v^>_{e,z})$ is the incident angle. For the Weyl node with a given chirality $\chi$, the $\theta_i$-resolved reflection is asymmetric, as shown in Fig.\ \ref{fig:4}(a), where the electrons with positive chirality ($\chi=+1$) have large reflection probabilities for $0^{\circ}<\theta_i<90^{\circ}$ (gray lines), whereas the electrons with negative chirality ($\chi=-1$) have large reflection probabilities for $-90^{\circ}<\theta_i<0^{\circ}$ (green lines). The carries with opposite chiralities turn into different transverse directions, leading to a transverse chirality current. With the help of Eqs.\ (\ref{Eq:app1}) and (\ref{Eq:app2}), the chirality Hall angle at $\alpha=0$ is obtained by
\begin{align}
\tan(\vartheta_\mathrm{chi})=2\sin\ell\times\frac{p\cos 2\ell+\cos \ell(\pi\sin\ell-2)}{\pi\cos2\ell-2p\sin 2\ell+4\sin\ell},
\end{align}
where $p=\mathrm{arctanh}(\cos\ell)$ with $\ell=\arccos(c)$. The chirality Hall angle is only determined by the tilting parameter $c$ and approaches its maximum value at $c\simeq0.82$ with $\vartheta_\mathrm{chi}\simeq0.22$, as shown in Fig.\ \ref{fig:4}(c) (black solid line). For the non-tilting energy dispersion ($c=0$), the transverse chirality current is absent.

Furthermore, it is found that the scattering pattern is mirror symmetric between two different chiral nodes, \textit{i.e.},
\begin{align}
R_{\chi,\alpha=0}(\theta_i)=R_{-\chi,\alpha=0}(-\theta_i),\label{Eq:aw0}
\end{align}
as shown in Fig.\ \ref{fig:4}(a), leading to a zero net transverse charge current with
\begin{align}
\tan(\vartheta)=0.
\end{align}
Consequently, the pure chirality current with zero net charge is predicted for $\alpha=0$.

For $\alpha=\pi$, the line connecting two opposite chiral Weyl nodes makes a $180^\circ$ rotation, implying the interchange of the chiralities between the nodes, which leads to the $\theta_i$-resolved $R_{\alpha=\pi}$ being a copy of $R_{\alpha=0}$ with the substitution $\chi\rightarrow-\chi$, as shown in Fig.\ \ref{fig:4}(b). Consequently, the pure transverse chirality Hall current is reversed, while its absolute value remains unchanged, as shown in Fig.\ \ref{fig:4}(c) (black dashed line).

\subsection{Symmetry analysis} 
\label{sub:3}

Distinct from the Berry curvature induced Hall effect \cite{mak2014valley,young2012spin,PhysRevX.9.031021}, the physical origin of the transverse Hall current in our model is attributed to the symmetry breaking caused by the the tilt. The low-energy Hamiltonian for the opposite chiral Weyl nodes at $\pm \mathbf{K}_0$ is given by $\mathcal{H}_\pm=\hbar v_F(\pm c q_t\sigma_0+q_x\sigma_x\mp q_y\sigma_y+ q_z\sigma_z)$. For the Weyl cones tilted in $q_x-q_z$ plane, the transverse Hall chirality current and charge current may occur when the angle-resolved reflection is asymmetric for a given chiral node, which requires the tilt breaking the mirror symmetry in the $q_y=0$ plane, \textit{i.e.},
\begin{align}
\mathcal{M}\mathcal{H}_{\pm}(q_x,q_z)\mathcal{M}^{-1}\neq\mathcal{H}_{\pm}(-q_x,q_z),\label{Eq:sm1}
\end{align}
where $\mathcal{M}=\sigma_z$ is the mirror reflection operator about the $y-z$ plane. For $q_t=q_x\cos\alpha-q_z\sin\alpha$, Eq.\ (\ref{Eq:sm1}) leads to $\cos\alpha\neq0$, \textit{i.e.}, $\alpha\neq\pm\pi/2$, which is in agreement with Eqs.\ (\ref{Eq:th1}-\ref{Eq:th2}).

Furthermore, in order to generate a pure transverse chirality Hall current, the reflection between two different chiral nodes should be symmetric to cancel out the net transverse charge current. This requires an additional $\mathbb{Z}_2$ exchange symmetry $\mathcal{C}$ \cite{PhysRevB.103.125147}, which swaps the opposite chiral sector: $\mathcal{C}\mathcal{H}_+(q_x,q_z)\mathcal{C}^{-1}=\mathcal{H}_-(q_x,q_z)$. Consequently, the transverse charge Hall current vanishes when the tilt preserves the combined $\mathcal{MC}$ symmetry
\begin{align}
(\mathcal{MC})\mathcal{H}_+(q_x,q_z)(\mathcal{MC})^{-1}=\mathcal{H}_-(-q_x,q_z),\label{Eq:sm2}
\end{align}
which results in $\alpha=0,\pi$.



\section{Conclusions} 
\label{sec:4}
To conclude, we study the transverse transport in the NS junctions based on the time-reversal symmetry broken type-I Weyl semimetals. We focus on the inversion symmetric tilt, where the two Weyl cones with opposite chiralities have tilts in opposite directions. Our investigation reveals that a chirality-contrasting skew reflection occurs at the NS interface due to the tilt of the Weyl cones, resulting in nonzero transverse chirality Hall currents. We further illustrate that the transverse chirality current here is determined by the symmetry of the tilt. Specifically, both transverse chirality current and charge current may occur when the tilt breaks the mirror symmetry ($\mathcal{M}$). However, a pure transverse chirality current with zero net charge is present when the tilt breaks $\mathcal{M}$ symmetry but preserves the combined $\mathcal{MC}$ symmetry with $\mathcal{C}$ being the $\mathbb{Z}_2$ exchange symmetry.

\onecolumngrid

\appendix

\section*{Appendix}

In this appendix we provide the details of the derivation of the transverse conductance and the Hall angle in Eqs.\ (\ref{Eq:app1}) and (\ref{Eq:app2}) of the main text.

The transverse current along the $\eta$ axis ($\eta=x,y$) is given by $I_\eta=I_\eta^>+I_\eta^<$ with $I^>_\eta$ ($I^<_\eta$) being the net current flowing from left to right (right to left) \cite{datta1997electronic,PhysRevLett.115.056602}. In order to get a balanced current in the barrier region, the state propagating towards (outwards) the barrier has a positive (negative) contribution to the transverse current. $I^>_\eta$ is carried by the state $|\varphi^>\rangle=|\varphi^>_e\rangle+r|\varphi^<_e\rangle+r_A|\varphi^<_h\rangle$ with energy $E$ and transverse wave vector $\mathbf{q}_\parallel$, which is given by

\begin{align}
I^>_\eta=&\frac{e}{L}\sum_{\mathbf{q}_\parallel,q_z}\Big(v_{e,\eta}^>-v_{e,\eta}^<|r|^2\Big)f(E-eV)[1-f(E)]\nonumber\\&-\left(-\frac{e}{L}\right)\sum_{\mathbf{q}_\parallel,q_z}v_{h,\eta}^<|r_A|^2[1-f(E+eV)]f(E)\nonumber\\
=&e\sum_{\mathbf{q}_\parallel}\int \frac{dq_z}{2\pi}\Big[\Big(v_{e,\eta}^>-v_{e,\eta}^<|r|^2\Big)f(E-eV)[1-f(E)]\nonumber\\&+v_{h,\eta}^<|r_A|^2[1-f(E+eV)]f(E)\Big]\nonumber\\
=&\frac{e}{h}\sum_{\mathbf{q}_\parallel}dE \Big[\Big(\frac{v_{e,\eta}^>}{v_{e,z}^>}-\frac{v_{e,\eta}^<}{v_{e,z}^>}|r|^2\Big)f(E-eV)[1-f(E)]\nonumber\\&+\frac{v_{h,\eta}^<}{v_{e,z}^>}|r_A|^2[1-f(E+eV)]f(E)\Big].
\end{align}
Here $V$ is the longitudinal voltage drop along the junction and $f(E)=1/(\mathrm{exp}(E/k_BT)+1)$ is the Fermi-Dirac distribution function with $k_B$ and $T$ being the Boltzmann constant and temperature, respectively.

The net current from right to left ($I^<_\eta$) can be obtained by considering the incoming states in the right superconducting region.  Instead of dealing with quasiparticles in the superconducting region, it is equivalent to suppose the incident hole state with energy $-E$ and transverse wave vector $-\mathbf{q}_\parallel$ in the left normal region, \textit{i.e.}, $|\psi^<\rangle=|\psi^>_h\rangle+\bar{r}|\psi^<_h\rangle+\bar{r}_A|\psi^<_e\rangle$, where $\bar r$ ($\bar r_A$) is the normal (Andreev) reflection amplitude for the hole state. Consequently, the net current carried by $|\psi^<\rangle$ reads
\begin{align}
I^<_\eta=&-\frac{e}{L}\sum_{\mathbf{q}_\parallel,q_z}\Big(v_{h,\eta}^>-v_{e,\eta}^<|\bar r|^2\Big)f(-E+eV)[1-f(-E)]\nonumber\\&-\frac{e}{L}\sum_{\mathbf{q}_\parallel,q_z}v_{e,\eta}^<|\bar r_A|^2[1-f(-E-eV)]f(-E)\nonumber\\
=&-\frac{e}{h}\sum_{\mathbf{q}_\parallel}\int dE \Big[\Big(\frac{v_{h,\eta}^>}{v_{h,z}^>}-\frac{v_{h,\eta}^<}{v_{h,z}^>}|\bar r|^2\Big)f(-E+eV)[1-f(-E)]\nonumber\\&+\frac{v_{e,\eta}^<}{v_{h,z}^>}|\bar r_A|^2[1-f(-E-eV)]f(-E)\Big].
\end{align}
The particle-hole symmetry leads to the follow identities
\begin{align}
\frac{v_{e,\eta}^>}{v_{e,z}^>}=&\frac{v_{h,\eta}^>}{v_{h,z}^>},\quad\frac{v_{e,\eta}^<}{v_{e,z}^>}|r|^2=\frac{v_{h,\eta}^<}{v_{h,z}^>}|\bar r|^2,\quad\frac{v_{h,\eta}^<}{v_{e,z}^>}|r_A|^2=\frac{v_{e,\eta}^<}{v_{h,z}^>}|\bar r_A|^2,
\end{align}

which simplify the expression for the total current 
\begin{align}
I_\eta=&I^>_\eta+I^<_\eta\nonumber\\
=&\frac{e}{h}\sum_{\mathbf{q}_\parallel}\int_{-\infty}^{\infty}dE \Big(\Big[\frac{v_{e,\eta}^>}{v_{e,z}^>}-\frac{v_{e,\eta}^<}{v_{e,z}^>}|r|^2\Big][f(E-eV)-f(E)]+\frac{v_{h,\eta}^<}{v_{e,z}^>}|r_A|^2[f(E)-f(E+eV)\Big)\nonumber\\
=&\frac{e}{h}\sum_{\mathbf{q}_\parallel}\int_{-\infty}^{\infty}dE \Big(\Big[\frac{v_{e,\eta}^>}{v_{e,z}^>}-\frac{v_{e,\eta}^<}{v_{e,z}^>}|r|^2\Big]_{(E)}+\Big[\frac{v_{h,\eta}^<}{v_{e,z}^>}|r_A|^2\Big]_{(-E)}\Big)[f(E-eV)-f(E)].
\end{align}
Consequently, the transverse conductance is given by
\begin{align}
\sigma_{\eta,z}=&\frac{\partial I_\eta}{\partial V}\nonumber\\
=&\frac{e^2}{h}\sum_{\mathbf{q}_\parallel}\int_{-\infty}^{\infty}dE\Big(\Big[\frac{v_{e,\eta}^>}{v_{e,z}^>}-\frac{v_{e,\eta}^<}{v_{e,z}^>}|r|^2\Big]_{(E)}+\Big[\frac{v_{h,\eta}^<}{v_{e,z}^>}|r_A|^2\Big]_{(-E)}\Big)\Big[-\frac{\partial(f(E-eV)-f(E))}{\partial (E-eV)}\Big]\nonumber\\
=&\frac{e^2}{h}\sum_{\mathbf{q}_\parallel}\int_{-\infty}^{\infty}dE\Big(\Big[\frac{v_{e,\eta}^>}{v_{e,z}^>}-\frac{v_{e,\eta}^<}{v_{e,z}^>}|r|^2\Big]_{(E)}+\Big[\frac{v_{h,\eta}^<}{v_{e,z}^>}|r_A|^2\Big]_{(-E)}\Big)\delta(E-eV)\nonumber\\
=&\frac{e^2}{h}\sum_{\mathbf{q}_\parallel}\Big(\Big[\frac{v_{e,\eta}^>}{v_{e,z}^>}-\frac{v_{e,\eta}^<}{v_{e,z}^>}|r|^2\Big]_{(eV)}+\Big[\frac{v_{h,\eta}^<}{v_{e,z}^>}|r_A|^2\Big]_{(-eV)}\Big).
\end{align}
The longitudinal current can be calculated by the similar method, which reads
\begin{align}
I_z=&I_z^>+I_z^<\nonumber\\
=&\frac{e}{h}\sum_{\mathbf{q}_\parallel}\int_{-\infty}^\infty dE \Big(\Big[1+\frac{v_{e,z}^<}{v_{e,z}^>}|r|^2\Big][f(E-eV)-f(E)]-\frac{v_{h,z}^<}{v_{e,z}^>}|r_A|^2[f(E)-f(E+eV)]\Big)\nonumber\\
=&\frac{e}{h}\sum_{\mathbf{q}_\parallel}\int_{-\infty}^\infty dE \Big(\Big[1+\frac{v_{e,z}^<}{v_{e,z}^>}|r|^2\Big]_{(E)}[f(E-eV)-f(E)]-\Big[\frac{v_{h,z}^<}{v_{e,z}^>}|r_A|^2\Big]_{(-E)}[f(E-eV)-f(E)]\Big).
\end{align}
Consequently, the longitudinal conductance is given by
\begin{align}
\sigma_{z,z}=&\frac{\partial I_z}{\partial V}\nonumber\\=&\frac{e^2}{h}\sum_{\mathbf{q}_\parallel}\int_{-\infty}^\infty dE\Big(\Big[1+\frac{v_{e,z}^<}{v_{e,z}^>}|r|^2\Big]_{(E)}-\Big[\frac{v_{h,z}^<}{v_{e,z}^>}|r_A|^2\Big]_{(-E)}\Big)\Big[-\frac{\partial(f(E-eV)-f(E))}{\partial (E-eV)}\Big]\nonumber\\
=&\frac{e^2}{h}\sum_{\mathbf{q}_\parallel}\int_{-\infty}^\infty dE\Big(\Big[1+\frac{v_{e,z}^<}{v_{e,z}^>}|r|^2\Big]_{(E)}-\Big[\frac{v_{h,z}^<}{v_{e,z}^>}|r_A|^2\Big]_{(-E)}\Big)\delta(E-eV)\nonumber\\
=&\frac{e^2}{h}\sum_{\mathbf{q}_\parallel}\Big(1+\Big[\frac{v_{e,z}^<}{v_{e,z}^>}|r|^2\Big]_{(eV)}-\Big[\frac{v_{h,z}^<}{v_{e,z}^>}|r_A|^2\Big]_{(-eV)}\Big).\label{btk}
\end{align}

It is noted that, for the incident (reflected) states, the longitudinal group velocities are always positive (negative), \textit{i.e.}, $v_{e,z}^><0$ and $v_{e/h,z}^<<0$. However, the sign of the transverse group velocity is uncertain. Consequently, for the longitudinal conductance, the reflection coefficients can be written as
\begin{gather}
\frac{v_{e,z}^<}{v_{e,z}^>}|r|^2=-\left|\frac{v_{e,z}^<}{v_{e,z}^>}\right||r|^2=-R,\\
\frac{v_{h,z}^<}{v_{e,z}^>}|r_A|^2=-\left|\frac{v_{h,z}^<}{v_{e,z}^>}\right||r_A|^2=-R_A.
\end{gather}
where $R_A$ ($R$) is positive and denotes the reflection probability for the Andreev reflection (normal reflection). In the subgap regime, $A+B=1$ due to the current conservation. The longitudinal conductance [Eq.\ (\ref{btk})] can be expressed in terms of $R$ and $R_A$
\begin{align}
\sigma_{z,z}=\frac{e^2}{h}\sum_{\mathbf{q}_\parallel}\Big(1-R+R_A\Big),
\end{align}
which is the Blonder-Tinkham-Klapwijk formalism. The charge Hall angle ($\vartheta$) and the chirality Hall angle ($\vartheta_{\mathrm{chi}}$) are given by
\begin{align}
    \tan(\vartheta)=\frac{\sum_\chi \sigma_{\eta,z}^\chi}{\sum_\chi\sigma_{z,z}^\chi},\quad  \tan(\vartheta_{\mathrm{chi}})=\frac{\sum_\chi \chi\sigma_{\eta,z}^\chi}{\sum_\chi\sigma_{z,z}^\chi}.
\end{align}

\twocolumngrid

\end{document}